\documentstyle[rotate,emulateapj,apjfonts]{article}

\slugcomment{Accepted by {\it The Astronomical Journal}}

\lefthead{THE X-RAY PROPERTIES OF $Z>4$ QUASARS}

\righthead{KASPI, BRANDT, \& SCHNEIDER}

\begin{document}

\title{The X-ray Properties of $z>4$ Quasars}
\author{Shai Kaspi, W. N. Brandt, and Donald P. Schneider}
\affil{Department of Astronomy and Astrophysics, 525 Davey Laboratory,
Pennsylvania State University, \\ University Park, PA, 16802;
shai, niel, dps@astro.psu.edu}

\begin{abstract}

We report on a search for X-ray emission from quasars with redshifts
greater than four using the {\it ROSAT} public database. Our search has
doubled the number of $z>4$ quasars detected in X-rays from 6 to 12.
Most of those known prior to this work were radio-loud and X-ray
selected sources; our study increases the number of X-ray detected,
optically selected $z>4$ quasars from one to seven.  We present their
basic X-ray properties and compare these to those of lower redshift
quasars. We do not find any evidence for strong broad-band spectral
differences between optically selected $z>4$ quasars and those at lower
redshifts.

\end{abstract}

\keywords{
galaxies: active --- 
galaxies: nuclei --- 
quasars: general ---
X-ray: galaxies}

\section{Introduction}

Quasars with redshifts larger than 4 were first discovered more than a
decade ago (Warren et~al. 1987). Recently many new $z>4$ quasars have
been discovered (e.g., Fan et~al. 1999, 2000) with expectations of an
even greater increase in the near future due to new surveys (e.g., the
Sloan Digital Sky Survey should identify $\approx 1000$ quasars with
$z>4.5$~-- Schneider 1999; York et~al. 2000). Currently there are 85
$z>4$ quasars that have appeared in journals, and an additional number
can be found on various World Wide Web pages. Quasars at $z>4$ provide
us with direct information about the first 10\% of cosmic time.  They
are among the most luminous objects known, and from the Eddington limit
many require $\ga$~$10^8$--$10^9$~M$_\odot$ black holes. They have wide
cosmological importance since they must be associated with deep
potential wells in the earliest massive collapsed structures, and their
strong evolution provides clues about the process by which the
remarkably homogeneous $z\approx 1000$ Universe revealed by the cosmic
microwave background is transformed into the inhomogeneous Universe
seen today (e.g., Efstathiou \& Rees 1988; Turner 1991).

To date $z>4$ quasars have been mainly studied at optical wavelengths
(e.g., in order to determine their redshifts). There has also been some
progress in studying their far-infrared and radio properties (e.g.,
Schmidt et~al. 1995; Omont et~al. 1996; McMahon et~al. 1999). However,
their properties as a whole at these and other wavelengths have not yet
been fully explored.  X-ray emission appears to be a universal property
of quasars at $z\approx$~0--2, and X-rays have also been studied from
many $z\approx$~2--4 quasars. However, at $z>4$ the X-ray properties of
quasars are much less well understood; only six $z>4$ quasars have been
detected in X-ray. The luminous X-ray emission from quasars reveals the
physical conditions in the immediate vicinities of their black holes,
and X-ray studies of high-redshift quasars can in principle discover if
quasar central power sources and quasar environments evolve over cosmic
time (e.g., Bechtold et~al. 1994b; Blair et~al. 1998; Elvis et~al.
1998; Fiore et~al. 1998).

We list the $z>4$ quasars previously detected in the X-ray band in
Table~1. For each quasar we give, in the first seven columns,
its coordinates, redshift, absorption-corrected X-ray flux in the
detection band, and the reference to the paper where the X-ray
detection was made. While the most basic X-ray properties (e.g.,
$\alpha_{\rm ox}$, the slope of a nominal power law between
2500~\AA\ and 2~keV) of these quasars appear to be generally consistent
with those of quasars at lower redshifts, the constraints are not tight
and require substantial improvement. Comparisons of these properties
with those of the majority of low-redshift quasars are difficult as
most of the X-ray detected $z>4$ quasars were selected in different
ways:  three of them are X-ray selected objects, two are radio
selected, and only one is optically selected.

All objects but one in Table~1 were detected using {\it
ROSAT}, demonstrating the ability of this satellite to detect $z>4$
quasars. Encouraged by this we have systematically searched for
detections of $z>4$ quasars in the {\it ROSAT} public database. In this
paper we present our results, which double the number of $z>4$ quasars
detected in X-rays. We increase the number of optically selected $z>4$
quasars from one to seven and provide limits for 15 others. In \S~2 we
present the database search and in \S~3 we discuss our results.

Throughout this paper we use the cosmological parameters $H_0=70$
km~s$^{-1}$~Mpc$^{-1}$ and $q_0=0.5$. We define the energy index
$\alpha$ as $f_\nu \propto \nu^{-\alpha}$ and likewise the photon index
$\Gamma = \alpha +1$ with photon flux density $f(E) \propto
E^{-\Gamma}$ in photons~cm$^{-2}$~s$^{-1}$~keV$^{-1}$. Unless otherwise
noted, we use $\alpha{_{\rm o}}=0.5$ in the UV-optical range and
$\alpha{_{\rm x}}=1$ in the X-ray range. These are representative
values of these parameters for lower redshift quasars (e.g., Netzer
1990; Reeves et~al. 1997).

\section{Search and Analysis}

We have searched the {\it ROSAT} public database\footnote{Via:
http://heasarc.gsfc.nasa.gov/W3Browse}
%
%
for all fields which include the optical positions of $z>4$ quasars in
the literature. We have found 27 quasars' positions (out of the total
85 $z>4$ quasars) to lie in {\it ROSAT} fields. For ten quasars we
found only one observation, while for the others there were two or
more. We retrieved from the {\it ROSAT} public database up to four
observations (when available) for each quasar. We preferentially chose
long observations where the quasar was close to the field's center.
Twenty-six quasars were observed by the Position Sensitive Proportional
Counter (PSPC; Pfeffermann et~al. 1987) with several of them also
having High Resolution Imager (HRI; David et~al. 1999) observations,
and one quasar was observed only with the HRI. Out of the 27 quasars,
12 were the observation's target and 15 were serendipitously in the
detector's field of view.

All observations were processed using the PROS software in
IRAF.\footnote{IRAF (Image Reduction and Analysis Facility) is
distributed by the National \\ Optical Astronomy Observatories, which are
operated by AURA, Inc., under coo- \\ perative agreement with the National
Science Foundation.} We have manually inspected the images (and
smoothed versions thereof) and have measured the net counts around the
optical positions of the quasars. Typically PSPC positions are good to
$\approx$20--30\arcsec\ (e.g., Voges et~al. 1996), and indeed all our
detections but one are $\la 20$\arcsec\ from the optical position. The
aperture size for count extraction was scaled to take into account the
size of the point spread function (PSF) at the off-axis angle of the
quasar. We were able to detect, at above the 2.5$\sigma$ level, an
X-ray source and extract the X-ray count rate for nine quasars. For the
other 18 quasars we were only able to determine upper limits to their
X-ray count rates. We have also searched for X-ray detections of 86
quasars listed on World Wide Web pages but not in the literature, and we
find none.

In the upper part of Table~\ref{properties} we present the quasars
detected. Their names, coordinates and redshifts are given in the first
four columns. The monochromatic AB$_{1450(1+z)}$ magnitude is listed in
column (5). The AB$_{1450(1+z)}$ magnitudes are from Schneider,
Schmidt, \& Gunn (1991), Henry et~al. (1994), Storrie-Lombardi et al.
(1996), and Hook \& McMahon (1998), with estimated errors of
$\approx\pm 0.1$ magnitudes. The Galactic column density, found using
the \ion{H}{1} map of Dickey \& Lockman (1990), is given in column (6).
We have used the AB$_{1450(1+z)}$ magnitude and a flux-density power
law with $\alpha_{\rm{0}} =0.5$ to compute the rest-frame 2500~\AA\ flux
density and luminosity which are listed in columns (7) and (8). The
absolute $B$ magnitude, given in column (9), was found using the
equation
\begin{equation}
M_B={\rm AB}_{1450(1+z)}-5\log(\sqrt{1+z}-1)-45.03
\end{equation}
which is equation~5 of Schneider, Schmidt, \& Gunn (1989) adapted for
$H_0=70$ km~s$^{-1}$~Mpc$^{-1}$.

In columns (10)--(18) of Table~\ref{properties} we list the X-ray
observations and properties of the quasars. The sequence-ID and
observing date are listed in columns (10) and (11). The angular
distance of the quasar's position from the center of the field is
listed in column (12). The number of background-subtracted counts (in
the broad band 0.1--2 keV) is given in column (13), and the vignetting,
exposure-map and background corrected count-rate is listed in column
(14). We use the PIMMS software (Mukai 1997) to define a power law with
photon index $\Gamma =2$, which is then used in the XSPEC software
(Arnaud 1996) to evaluate the absorption-corrected 0.1--2 keV observed
flux, which is listed in column (15). From this power law we also
calculated the rest-frame 2-keV flux density and luminosity which are
listed in columns (16) and (17). (Using PROS to calculate these
quantities yields consistent results.) Finally, we list in column (18)
the effective optical-to-X-ray power-law slope, $\alpha_{\rm ox}$,
defined as:
\begin{equation}
\alpha_{\rm ox}=\frac{\log[(f_\nu (2\,{\rm keV})/f_\nu (2500\,{\mbox\AA})]}
{\log[\nu (2\,{\rm keV})/\nu (2500\,{\mbox\AA})]} \ \ , 
\end{equation}
where $f_\nu$'s are flux densities at the given wavelengths and $\nu$'s
are the corresponding frequencies.  The uncertainties for the fluxes,
luminosities, and $\alpha_{\rm ox}$ values can be propagated from the
relative error on the number of counts, which has a mean value of
30\%.  Another uncertainty involves our assumption of $\Gamma =2$. 

%
\footnotesize
\rotate[l]{\makebox[0.95\textheight][l]{\vbox{
\begin{center}
\centerline{\sc \hglue 6in Table~1}
\centerline{\sc \hglue 5.5in X-ray Detected Quasars at $z>4$ Known Prior to This Work}
\vglue 0.1in
\begin{tabular}{lcclcccccccccccr}\hline\hline
{Object} & \multicolumn{2}{c}{RA \ \ (2000.0) \ \ Dec} &
{$z$}           & {Flux\tablenotemark{a}}  & {Band [keV]}  & {Ref.}  &
{AB}            & {$N_{\rm H}$\tablenotemark{b}}  &
{$f_{\nu}$\tablenotemark{c}}  & {$\log (\nu L_\nu )$}  & {$M_B$} &
{$f_\nu$\tablenotemark{d}} & {$\log (\nu L_\nu )$}  & {$\alpha{_{\rm
ox}}$} & {$R$} \\
{(1)} & {(2)} & {(3)} & {(4)} & {(5)} & {(6)} & {(7)} & {(8)} & {(9)} &
{(10)} & {(11)} & {(12)} & {(13)} & {(14)} & {(15)} & {(16)}     \\
\hline
Q\,0000$-$2619     & 00 03 22.9  & $-$26 03 19  &  4.098 &  8.7$\times10^{-14}$\,\tablenotemark{e}  & 0.1--2.4  & 1    &
17.5 & 1.67 & 4.767 & 46.9 & $-$28.0 & 2.9 & 45.3 & $-$1.62 & $<$0.6 \\
RX\,J1028.6$-$0844 & 10 28 37.7  & $-$08 44 39  &  4.276 &  8.3$\times10^{-13}$  & 0.1--2.4  & 2    &
20.6\tablenotemark{f}& 4.55 & 0.274 & 45.7 & $-$25.0 & 28.6& 46.3 & $-$0.76 & 7400 \\
RX\,J105225.9+571905 & 10 52 25.9 &  +57 19 07 & 4.45 &  2.3$\times10^{-15}$  & 0.5--2.0  & 3    &
22.6&0.56& 0.043 & 44.9 & $-$23.1 & 0.19& 44.2 & $-$1.29 & $<$190 \\
GB\,1428+4217 &  14 30 23.7 & +42 04 36 & 4.715 & $\sim1\times10^{-12}$  & 0.1--2.4  & 4    &
19.4 & 1.40 & 0.829 & 46.2 & $-$26.4 & 37.2& 46.5 & $-$0.90 & 1900 \\
GB\,1508+5714 & 15 10 02.8 & +57 02 44 & 4.301 & $1\times10^{-12}$  & 0.3--3.5  & 5, 6    &
19.8 & 1.47 & 0.573 & 46.0 & $-$25.8 & 44.7& 46.5 & $-$0.81 & 2700 \\
RX\,J1759.4+6638 & 17 59 27.9 & +66 38 53 & 4.320 & $\sim1.2\times10^{-14}$  & 0.5--2.0  & 7    &
19.3 & 4.23 & 0.908 & 46.2 & $-$26.3 & 0.95& 44.8 & $-$1.53 & 30 \\
\hline
\end{tabular}
\end{center}
$^{\rm a}${In units of erg\,s$^{-1}$\,cm$^{-2}$.}\\
$^{\rm b}${In units of $10^{20}$ cm$^{-2}$.}\\
$^{\rm c}${At 2500~\AA\ in units of $10^{-27}$ ergs\,s$^{-1}$\,cm$^{-2}$\,Hz$^{-1}$.}\\
$^{\rm d}${At 2~keV in units of $10^{-31}$ ergs\,s$^{-1}$\,cm$^{-2}$\,Hz$^{-1}$.}\\
$^{\rm e}${Computed using Bechtold et al. (1994a) model \#~3.} \\
$^{\rm f}${Estimated from the 2500~\AA\ flux given in
Zickgraf et al. (1997).} \\
\vglue 0.01in
\centerline{ \hglue 4.75in {\sc References.---}
(1) Bechtold et al. 1994a; 
(2) Zickgraf et al. 1997;  
(3) Schneider et al. 1998; 
(4) Fabian et al. 1997;    
(5) Mathur \& Elvis 1995;  
(6) Moran et~al. 1996;     
(7) Henry et al. 1994.   } 
}}}
\setcounter{table}{1}
\normalsize

%
\scriptsize
\begin{table*}
\tablenum{2}
\rotate[l]{\makebox[0.95\textheight][l]{\vbox{
\begin{center}
{\sc Table~2 \\ Properties of Quasars at $z>4$ Observed by {\it ROSAT}}
\vglue 0.1in
\begin{tabular}{lcclccccccccccccccr}
\hline
\hline
{Object} & \multicolumn{2}{c}{RA \ \ (2000.0) \ \ Dec} &
{$z$}           & {AB}            & {$N_{\rm H}$\tablenotemark{a}}  &
{$f_{\nu}$\tablenotemark{b}}  & {$\log (\nu L_\nu )$}  & {$M_B$}    &
{Sequence-ID}   & {Date}          & {$\Delta$\tablenotemark{c}}     &
{Counts}        & {C-rate\tablenotemark{d}}       &
{$f$\tablenotemark{e}}                  & {$f_\nu$\tablenotemark{f}} &
{$\log (\nu L_\nu )$}  & {$\alpha_{\rm ox}$} & {$R$} \\ [0.1cm]
{(1)} & {(2)} & {(3)} & {(4)} & {(5)} & {(6)} & {(7)} & {(8)} & {(9)} &
{(10)} & {(11)} & {(12)} & {(13)} & {(14)} & {(15)} & {(16)} & {(17)} &
{(18)} & {(19)} \\ [0.1cm]
\hline
\multicolumn{19}{c}{Detected Quasars} \\ [0.1cm]
\hline
Q\,0000$-$2619 & 00 03 22.9 & $-$26 03 19 & 4.098 & 17.5 & 1.67 & 
4.767 & 46.9 & $-$28.0 & 
rp700467n00 & 26/11/91 & 0.6 &
177$\pm$24 & 4.97 & 6.5 & 2.4 & 45.2 & $-$1.65 & $<$0.6 \\ [0.1cm]
 &  &  &  &  &  & 
 &  &  & 
rp700078n00 & 30/11/91 &  0.4 &
30.0$\pm$9.3 & 7.29 & 9.5 & 3.5 & 45.4 & $-$1.59 & \\ [0.1cm]
BR\,0019$-$1522 & 00 22 08.0 & $-$15 05 39 & 4.528 & 18.8 & 2.09 &
1.440 & 46.4 & $-$26.9 &
rp701207n00 & 23/06/92 & 0.1 &
34$\pm$11 & 5.36 & 7.9 & 3.1 & 45.4 & $-$1.41 & $<$2.0 \\ [0.1cm]
 &  &  &  &  &  & 
 &  &  & 
rp701207a01 & 08/12/92 & 0.1 &  
16.8$\pm$6.3 & 4.87 & 7.2 & 2.8 & 45.3 & $-$1.42 & \\ [0.1cm]
BR\,0351$-$1034 & 03 53 46.9 & $-$10 25 19 & 4.351 & 18.7\tablenotemark{g} & 4.08 &
1.579 & 46.5 & $-$26.9 &
rp700531n00 & 27/01/92 & 0.4 &
54$\pm$13 &  5.92 & 12.4 & 4.7 & 45.5 & $-$1.35 & $<$3.1 \\ [0.1cm]
BR\,0951$-$0450 & 09 53 55.7 & $-$05 04 19 & 4.369 & 19.2 & 3.78 &
0.996 & 46.3 & $-$26.4 &
rp700379n00 & 18/05/92 & 0.0 & 
35$\pm$10 & 4.25 & 8.6 & 3.3 & 45.4 & $-$1.34 & $<$3.4 \\ [0.1cm]
BRI\,0952$-$0115 & 09 55 00.1 & $-$01 30 07 & 4.426 & 18.7 & 3.96 &
1.579 & 46.5 &  $-$27.0 &
rp700380n00 & 29/05/92 & 0.1 &
18.9$\pm$7.7 & 5.63 & 11.6 & 4.5 & 45.5 & $-$1.36 & $<$2.1 \\ [0.1cm]
BR\,1202$-$0725 & 12 05 23.1 & $-$07 42 32 & 4.695 & 18.0 & 3.46 &
3.008 & 46.8 & $-$27.7 &
rp700530n00 & 13/12/91 & 0.4 &
34$\pm$13 & 3.56 & 6.8 & 2.8 & 45.4 & $-$1.55 & $<$2.0 \\ [0.1cm]
GB\,1428+4217 & 14 30 23.7 & $+$42 04 36 & 4.715 & 19.4 & 1.40 &
0.829 & 46.2 & $-$26.4 &
rh704036n00 & 11/12/97 & 0.2 &
75$\pm$13 & 15.4 & 74.1 & 27.5 &  46.4 & $-$0.95 & 1900 \\ [0.1cm]
 &  &  &  &  &  & 
 &  &  & 
rh704007n00 & 09/01/98 & 0.2 &
125$\pm$17 & 21.5 & 102.9 & 38.3 & 46.5 & $-$0.90 & \\ [0.1cm]
 &  &  &  &  &  & 
 &  &  & 
rh704008n00 & 22/01/98 & 0.2 &
223$\pm$20 & 36.1 & 173.3 & 64.4 & 46.7 & $-$0.81 & \\ [0.1cm]
RX\,J1759.4+6638 & 17 59 27.9 & +66 38 53 & 4.320 & 19.3 & 4.23 &
0.908 & 46.2 & $-$26.3 &
rp000026n00 & 21/02/92 & 6.3 &
82$\pm$29 & 1.94 & 4.1 & 1.6 & 45.1 & $-$1.45 & 30 \\ [0.1cm]
BR\,2237$-$0607 & 22 39 53.6 & $-$05 52 19 & 4.558 & 18.3\tablenotemark{g} & 3.84 &
2.282 & 46.7 & $-$27.4 &
rp701206n00 & 20/05/93 & 0.1 &
56$\pm$13 & 6.17 & 12.5 & 4.9 & 45.6 & $-$1.41 & $<$3.6 \\ [0.1cm]
 &  &  &  &  &  & 
 &  &  & 
rh800789n00 & 26/05/96 & 9.1 &
30$\pm$17 & 1.30 & 8.4 & 3.0 & 45.4 & $-$1.45 & \\ [0.1cm]
\hline
\multicolumn{19}{c}{Undetected Quasars -- X-ray Upper Limits} \\  [0.1cm]
\hline
PC\,0027+0525          & 00 29 49.9 & +05 42 04 & 4.099 & 21.4 & 3.42 & 
0.131 & 45.3 & $-$24.1 &
rp201077n00 & 09/07/92 & 50.1 & 
 66.2 & 15.13 & 28.9 & 10.5 & 45.8 & $-$0.81 & 280 \\  [0.1cm]
PC\,0027+0521          & 00 30 04.6 & +05 38 13 & 4.210 & 22.3 & 3.43 & 
0.057 & 45.0 & $-$23.3 &
rp201077n00 & 09/07/92 & 49.8 & 
 80.7 & 18.41 & 35.2 & 13.0 & 46.0 & $-$0.63 & $<$220 \\  [0.1cm]
Q\,0046$-$293           & 00 48 29.6 & $-$29 03 21 & 4.014 & 19.3 & 1.75 & 
0.908 & 46.2 & $-$26.2 &
rp700275n00 & 01/06/92 & 47.2 & 
134.0 & 10.00 & 13.4 &  4.8 & 45.5 & $-$1.26 & $<$10 \\  [0.1cm]
Q\,0051$-$279          & 00 54 15.4 & $-$27 42 08 & 4.395 & 19.2 & 1.72 & 
0.996 & 46.3 & $-$26.4 &
rp701223n00 & 03/07/92 & 43.1 & 
133.8 &  4.38 &  5.8 &  2.2 & 45.2 & $-$1.40 & $<$4.9 \\  [0.1cm]
Q\,0101$-$304           & 01 03 37.3 & $-$30 08 59 & 4.072 & 20.0 & 2.03 & 
0.477 & 45.9 & $-$25.5 &
rp701194n00 & 06/07/92 &  0.4 & 
 18.5 &  1.38 &  2.0 &  0.7 & 44.7 & $-$1.47 & $<$14 \\  [0.1cm]
BRI\,0103+0032          & 01 06 19.2 & +00 48 23 & 4.433 & 18.8 & 3.19 & 
1.440 & 46.4 & $-$26.7 &
rh703871n00 & 22/12/97 & 17.1 & 
111.4 &  8.89 & 54.2 & 19.2 & 46.2 & $-$1.10 & $<$3.7 \\  [0.1cm]
SDSS033829.31+002156.3 & 03 38 29.3 & +00 21 56 & 5.000 & 20.0 & 8.11 & 
0.477 & 46.0 & $-$25.8 &
rp200844n00 & 26/01/92 & 28.5 & 
 26.2 & 12.70 & 35.6 & 15.2 & 46.1 & $-$0.96 & $<$9.6 \\  [0.1cm]
PC\,0953+4749\tablenotemark{h} & 09 56 25.2 & +47 34 44 & 4.457 & 19.1 & 0.98 & 
1.092 & 46.3 & $-$26.6 &
rp700450n00 & 12/05/92 & 13.5 & 
 21.4 &  6.53 &  6.6 &  2.5 & 45.3 & $-$1.39 & $<$4.1 \\ [0.1cm]
RX\,J105225.9+571905\tablenotemark{i}    & 10 52 25.9 & +57 19 07 & 4.450 & 22.6 & 0.56 & 
0.043 & 44.9 & $-$23.1 &
rp900029a00 & 16/04/91 &  4.0 & 
 60.5 &  0.97 &  0.8 &  0.3 & 44.4 & $-$1.21 & $<$190 \\  [0.1cm]
BRI\,1050$-$0000          & 10 53 20.4 & $-$00 16 49 & 4.291 & 19.4 & 3.95 & 
0.828 & 46.2 & $-$26.2 &
rp700381n00 & 26/05/92 &  0.4 & 
 18.2 &  4.00 &  8.2 &  3.1 & 45.3 & $-$1.32 & 88 \\  [0.1cm]
BR\,1144$-$0723         & 11 46 35.6 & $-$07 40 05 & 4.147 & 18.8 & 3.75 & 
1.440 & 46.4 & $-$26.7 &
rp700382n00 & 05/06/92 &  0.3 & 
 13.4 &  2.94 &  5.9 &  2.2 & 45.2 & $-$1.47 & $<$8.8 \\  [0.1cm]
SDSS122600.68+005923.6  & 12 26 00.7 & +00 59 24 & 4.250 & 19.1 & 1.89 & 
1.092 & 46.3 & $-$26.5 &
rp600242a01 & 24/12/92 & 44.7 & 
131.6 & 12.61 & 17.6 &  6.5 & 45.7 & $-$1.24 & $<$3.4 \\  [0.1cm]
PC\,1233+4752          & 12 35 31.1 & +47 36 06 & 4.447 & 20.1 & 1.18 & 
0.435 & 45.9 & $-$25.6 &
rp200578n00 & 14/11/91 & 22.8 & 
 33.7 & 21.07 & 23.1 &  8.9 & 45.8 & $-$1.03 & $<$21 \\  [0.1cm]
PKS\,1251$-$407           & 12 53 59.5 & $-$40 59 31 & 4.458 & 19.6\tablenotemark{j} & 7.97 & 
0.689 & 46.1 & $-$26.1 &
rp800321a01 & 18/01/93 & 36.4 & 
 59.6 &  9.27 & 25.8 & 10.0 & 45.9 &  $-$1.09 & 2800\tablenotemark{j} \\ [0.1cm]
SDSS131052.52$-$005533.4  & 13 10 52.5 & $-$00 55 33 & 4.140 & 18.9 & 1.77 & 
1.313 & 46.3 & $-$26.6 &
rp800248n00 & 18/07/92 & 26.8 & 
 49.6 &  6.51 &  8.8 &  3.2 & 45.3 & $-$1.39 & $<$0.7 \\  [0.1cm]
Q\,2133$-$4311          & 21 36 23.7 & $-$42 58 18 & 4.200 & 20.9\tablenotemark{g,k} & 2.84 &
0.208 & 45.6 & $-$24.7 &
rp800336a01 & 06/05/93 & 30.8 & 
 43.6 &  5.35 &  9.3 &  3.5 & 45.4 &  $-$1.06 & \nodata \\ [0.1cm]
Q\,2139$-$4324          & 21 42 58.2 & $-$43 10 59 & 4.460 & 20.7\tablenotemark{g,k} & 2.45 & 
0.250 & 45.7 & $-$25.0 &
rp300274n00 & 28/04/93 & 29.7 & 
 65.3 & 11.69 & 18.7 &  7.3 & 45.7 &  $-$0.97 & \nodata \\ [0.1cm]
PC\,2331+0216          & 23 34 31.9 & +02 33 22 & 4.093 & 19.8 & 4.81 & 
0.573 & 46.0 & $-$25.7 &
rh701901n00 & 24/12/94 &  0.2 & 
 30.1 &  1.93 & 13.4 &  4.4 & 45.5 & $-$1.19 & 35  \\  [0.1cm]
\hline
\end{tabular}
\end{center}
$^{\rm a}${In units of $10^{20}$ cm$^{-2}$.} \\
$^{\rm b}${At 2500~\AA\ in units of $10^{-27}$ ergs\,s$^{-1}$\,cm$^{-2}$\,Hz$^{-1}$.} \\
$^{\rm c}${Offset from the field center in units of arcmin.} \\
$^{\rm d}${In units of $10^{-3}$  counts\,s$^{-1}$.}\\
$^{\rm e}${Flux at the 0.1--2 keV band in units of $10^{-14}$ ergs\,s$^{-1}$\,cm$^{-2}$.} \\
$^{\rm f}${At 2~keV in units of $10^{-31}$ ergs\,s$^{-1}$\,cm$^{-2}$\,Hz$^{-1}$.} \\
$^{\rm g}${Estimated using an empirical linear relation between
[APM\,$R - {\rm AB}_{1450(1+z)}$] and [$z$] for all objects in
Storrie-Lombardi (1996) and deriving the missing AB$_{1450(1+z)}$
from that relation.} \\
$^{\rm h}${For better upper-limit determination see Molthagen,
Wendker, \& Briel (1995).} \\
$^{\rm i}${Can be only detected by careful co-adding of many
observations (see Table~1 and Schneider et al. 1998).} \\
$^{\rm j}${From Shaver et~al. (1996); using the $m_i$ magnitude
to estimate AB and the 1.4 GHz flux density to estimate $R$.} \\
$^{\rm k}${Estimated using $m_R$ magnitudes from and Hawkins \& 
V\'{e}ron (1996).}
}}}
\label{properties}
\end{table*}
\normalsize

\begin{figure*}
\vspace{-4cm}
\centerline{\epsfxsize=20.0cm\epsfbox{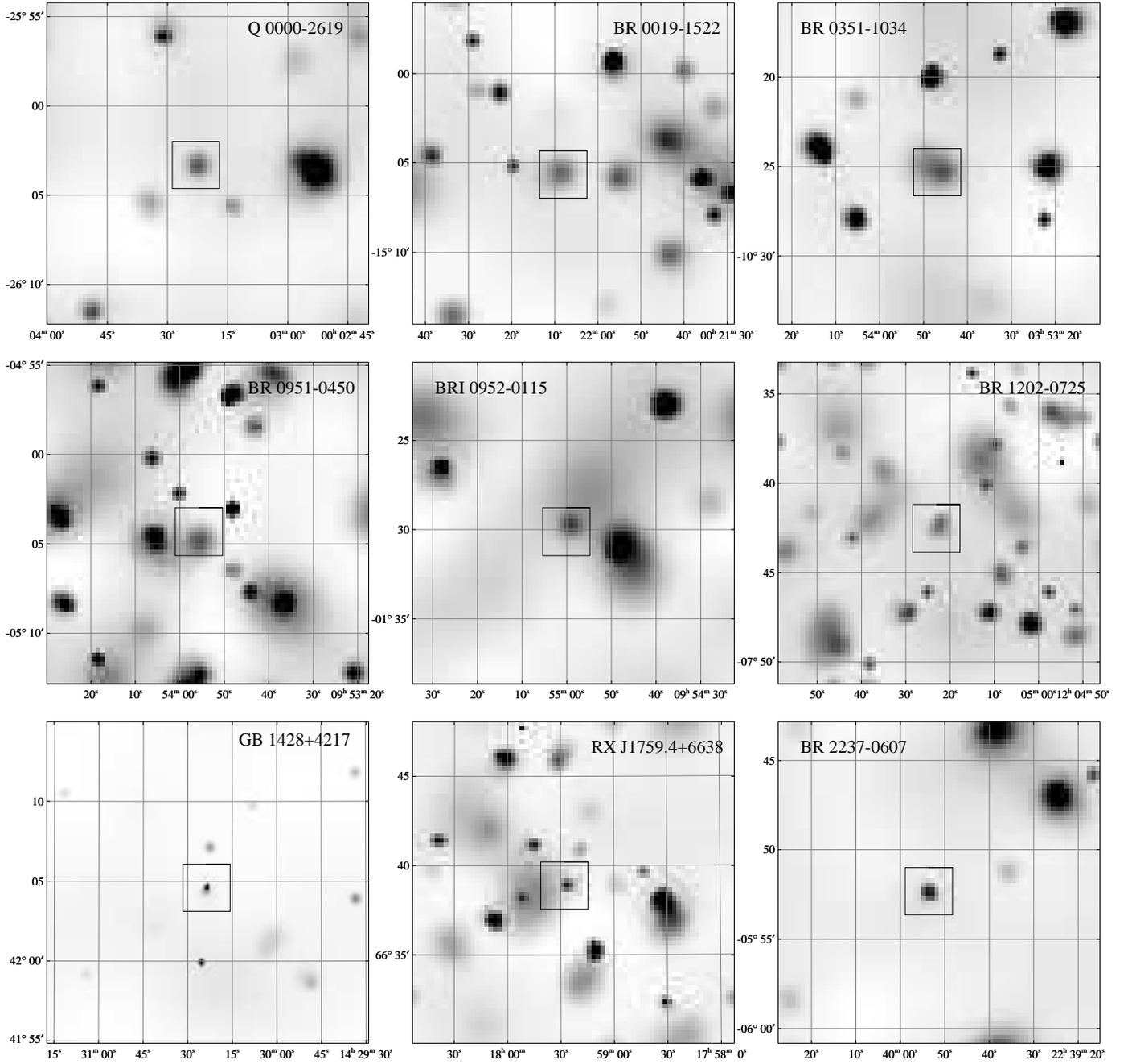}}
\vspace{-6cm}
\caption{{\it ROSAT} images of the nine detected $z>4$ quasars from
0.1--2.4 keV. North is up and East is to the left. The horizontal axis
shows the Right-Ascension, and the vertical axis shows the Declination
in J2000 coordinates. The box in each image is centered on the optical
coordinates of the quasar, and its size is substantially larger than
the positions uncertainty. Each image is
$\approx$18\arcmin$\times$18\arcmin . All images are from the PSPC
except for GB\,1428+4217 which is from the HRI.}
\label{image}
\end{figure*}

\noindent As radio-quiet quasars typically have $\Gamma$\,=\,1.7--2.3, our
assumption of $\Gamma =2$ might introduce an additional uncertainty of
$\sim$15\% in the flux estimates. As the true $\Gamma$ for each quasar
is unknown (other than GB\,1428+4217 and GB\,1508+5714), we do not
quote an error on the latter quantities. We estimate the total
fractional uncertainties for the fluxes to be in the range of
30--50\%.

\vglue 0.06cm
X-ray images of all nine detected quasars are presented in
Fig.~\ref{image} which was created using the adaptive smoothing method
of Ebeling, White, \& Rangarajan (2000) applied to the full-band
images. For most of the images we used a 2.5$\sigma$ level of
smoothing; for three low-$\sigma$ detected objects (BR\,0951$-$0450,
BRI0952$-$0115, and BR\,1202$-$0725) we used a 2.0$\sigma$ level of
smoothing.

\vglue 0.06cm
For each of the PSPC observations listed in the upper part of
Table~\ref{properties} we also calculated the X-ray counts in the soft
band (0.1--0.5 keV) and the hard band (0.5--2.0 keV). We used these to
calculate the rest-frame 2-keV flux density, luminosity, and
$\alpha_{\rm ox}$, in the same way as described above. The results were
similar to the broad-band results.  However, as the statistical
uncertainties in the soft and hard bands were large due the small
number of counts (note the small number of counts in the broad band --
column [13] of Table~\ref{properties}), we do not include these results
in our analysis.

\vglue 0.06cm
In the lower part of Table~\ref{properties} we list the $z>4$ quasars
which we were unable to detect in the X-ray observations.  For these
quasars we give 3$\sigma$ upper limits (listed in column [13]), where
$\sigma$ is the square root of the counts in an aperture of size
appropriate to the quasar's off-axis angle and centered at the quasar's
optical position. For each quasar we list the one observation which
gave the faintest upper limit.

\vglue 0.06cm
For 27 objects the number of expected detections which are merely
statistical fluctuations at the 2.5$\sigma$ level is 0.17; this
suggests that none of our detections is likely to be a background
fluctuation. We also estimated the probability that our X-ray quasar
detections are merely of unrelated X-ray sources that happen to be
coincident with the $z>4$ quasars' optical positions. We shifted the 27
quasars' positions by eight arcmin in eight different directions
and looked again for X-ray detections (in the same manner as was done
for the real optical positions). At the new positions we found seven
X-ray sources which met our detection criteria regarding significant
level and positional coincidence. This test shows that among our nine
X-ray detections there might be one source which is not the counterpart
of the quasar but an X-ray source which happened to be in that position
by chance. Additional suggestive evidence that the number of false
detections is small is that we have detected only the brighter quasars
out of the total 27 (see Fig.~\ref{his} and \S~\ref{Discussion}); if our
detections had random contamination by foreground sources then 
their distribution would not be correlated with quasar luminosity.

\vglue 0.06cm
For comparison purposes we present some properties of the X-ray
detected $z>4$ quasars known prior to this work in Table~1.
We list the AB$_{1450(1+z)}$ magnitudes in column (8) and the Galactic
column densities in column (9). Based on the AB$_{1450(1+z)}$
magnitudes we calculate the rest-frame 2500~\AA\ flux densities and
luminosities which are listed in columns (10) and (11), and the
absolute $B$ magnitudes in column (12). We used the X-ray fluxes and
bands from columns (5) and (6) and the PIMMS software to define a
simple power law with $\Gamma =2$. This power law was used to estimate
the rest-frame 2-keV flux den- 
%
\centerline{\epsfxsize=8.5cm\epsfbox{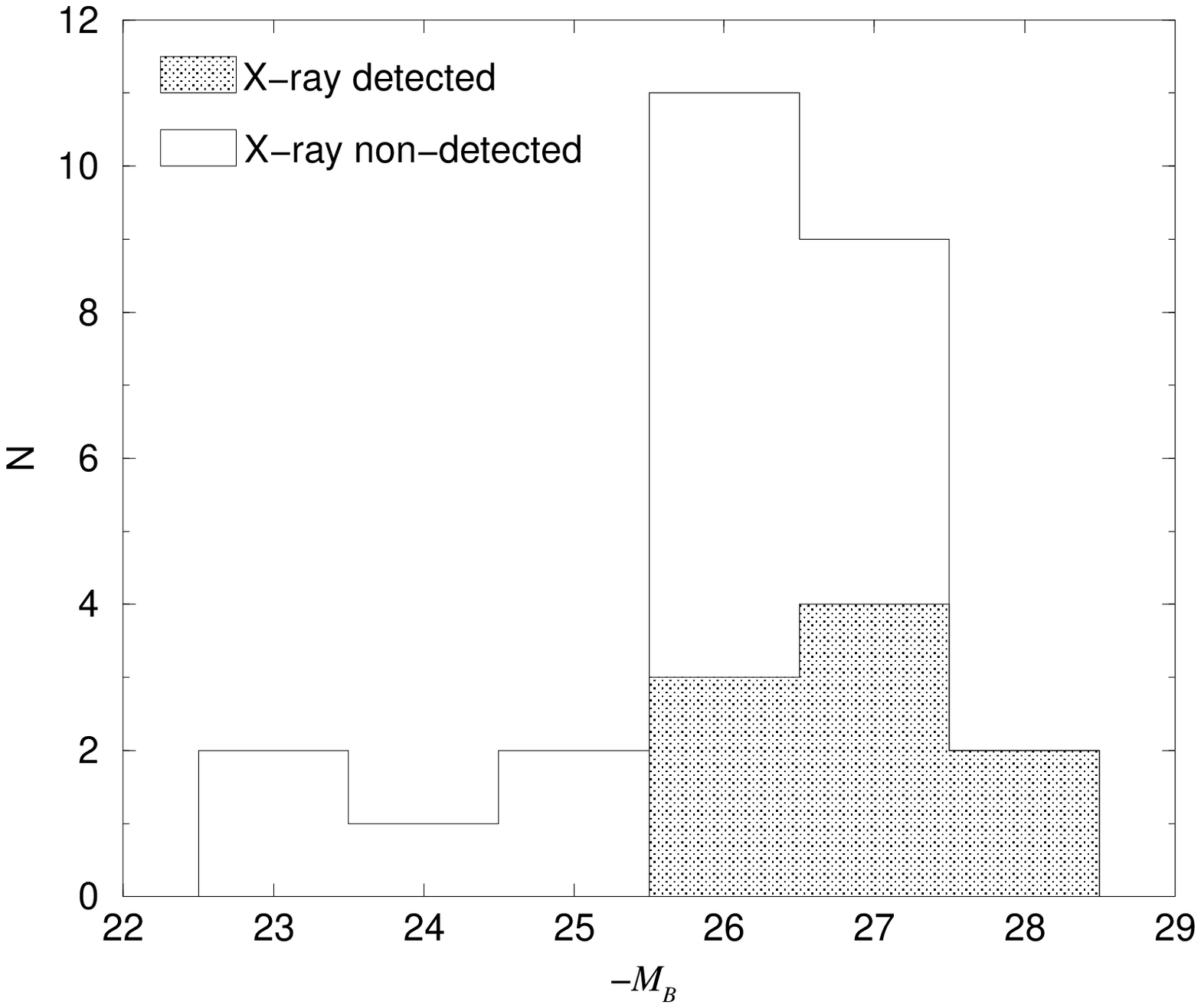}}
\figcaption{$M_B$ distribution of the 27 $z>4$ quasars observed by {\it
ROSAT}. The shaded area represents the quasars which were X-ray
detected. \label{his} }
\centerline{}
\centerline{}

\noindent sity and luminosity which are listed in
columns (13) and (14). We list the resulting $\alpha_{\rm ox}$ in column (15).

\section{Discussion}
\label{Discussion}

In most cases where we detect X-ray sources at the quasars' optical
positions they are very close to the detection limit.  Comparing the
data for the quasars which are X-ray detected to those which are not,
we notice several trends. The detected quasars are among the brighter
in the optical band (see Fig.~\ref{his} for the $M_B$ distribution). If
the undetected quasars have comparable X-ray luminosities to the
detected ones, we suggest that they were mainly not detected since they
were observed at the edges of the PSPC field where the PSF and
vignetting are larger. In the two cases where high luminosity
quasars were observed close to the PSPC field center (BRI\,1050$-$0000,
BR\,1144$-$0723), we attribute the non-detections to short exposure
times.

Three objects which we detected had already been reported as X-ray
emitters in the past (Q\,0000$-$2619, GB\,1428+4217, and
RX\,J1759.4+6638). The X-ray properties we have measured for them are
in agreement with the previous reported properties (see
Table~1). We have not detected the other three quasars in
Table~1 since the data for RX\,J1028.6$-$0844 are not public
(it was detected in the {\it ROSAT} All-Sky Survey), GB\,1508+5714 was
not observed by {\it ROSAT}, and RX\,J105225.9+571905 could be detected
only in the ``ultradeep'' HRI survey (see Schneider et~al. 1998 and
references therein).

At present only a few radio-loud $z>4$ quasars are known. These include
GB\,1428+4217, GB\,1508+5714, and RX\,J1028.6$-$0844 which have had
their X-ray data published prior to this work (Table~1),
PKS\,1251$-$407, for which we present an X-ray upper limit, and
GB\,1713+2148, whose {\it ROSAT} observation is not yet public. Out of
these objects, GB\,1428+4217 and GB\,1508+5714 show evidence for
relativistic beaming (e.g., Moran \& Helfand 1997; Fabian et~al.
1999). We used the NRAO\footnote{The National Radio Astronomy
Observatory\,is\,a\,facility\,of\,the National Science Foundation operated
under cooperative  agreement\,by\,Associated Universities,\,Inc.} VLA Sky
Survey (NVSS; Condon et~al. 1998) catalog and images at 1.4~GHz to
estimate the quasars' radio loudnesses, $R$, defined as the ratio of
the radio flux (extra-

\centerline{\epsfxsize=8.5cm\epsfbox{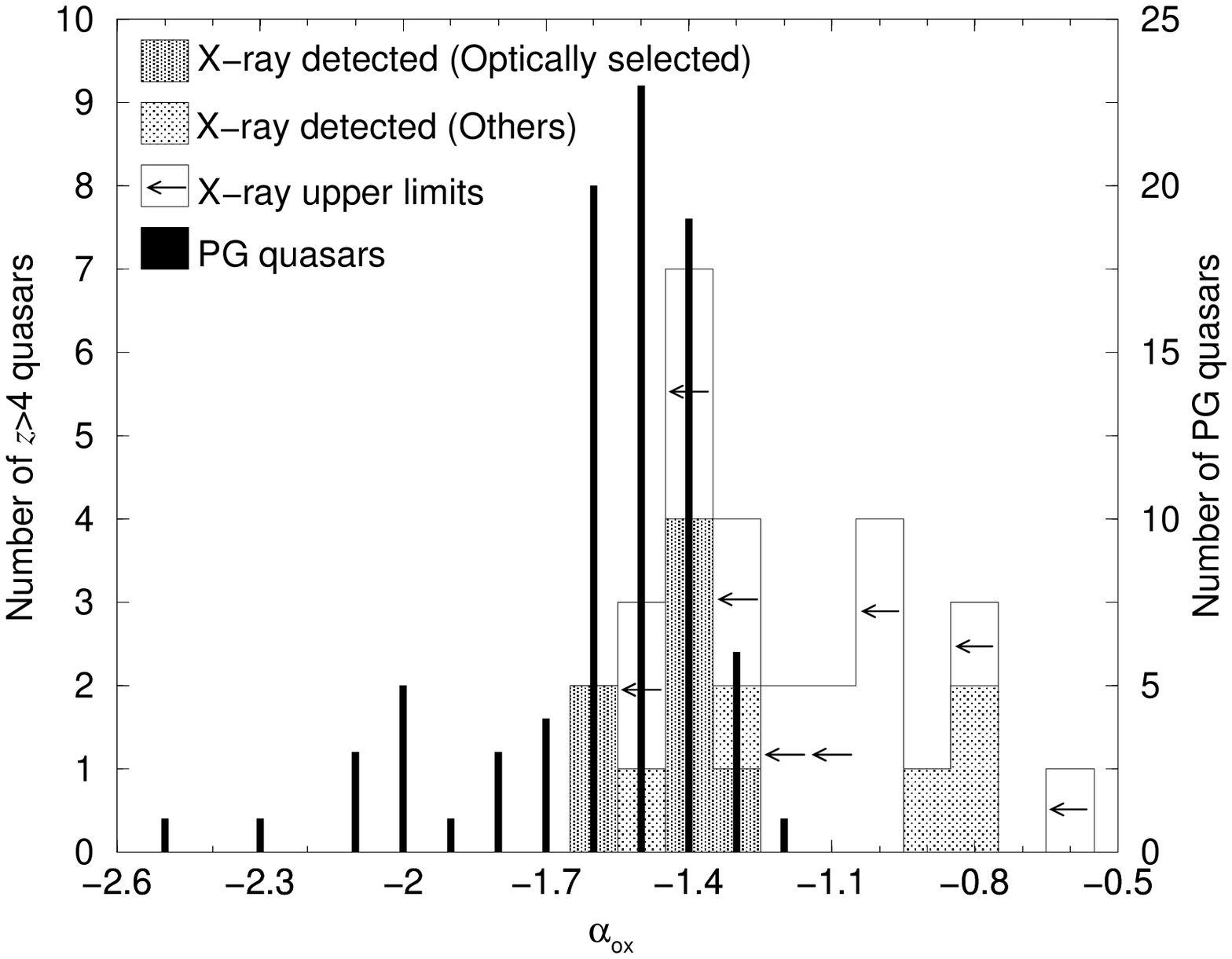}}
\vglue 0.37cm
\figcaption{$\alpha_{\rm ox}$ distribution of X-ray detected $z>4$ quasars
(shaded areas) compared with the $\alpha_{\rm ox}$ distribution of the 87
$z<0.5$ PG quasars (black bars) from Brandt et~al. (2000; including
four upper limits). We distinguish between $z>4$ quasars which are
optically selected and those which are selected from X-ray and radio
samples. The distribution of $z>4$ X-ray upper-limits is also shown
(white bars with arrows).
\label{aoxhis}}
\centerline{}
\centerline{}

\noindent cted from the NVSS) to the optical flux at
4400~\AA\ (estimated using the AB magnitude and a flux-density power
law with $\alpha_{\rm{o}}=0.5$). Radio-loud quasars typically have
$R>100$ and radio-quiet quasars have $R<10$ (e.g., Kellermann et~al.
1989). We list $R$ in Table~1 column (16) and
Table~\ref{properties} column (19). Most of the quasars in this study
are undetected by the NVSS, and thus we provide only upper
limits\footnote{The results reported here from the NVSS are consistent
with those from the Faint Images of the Radio Sky at Twenty-cm (FIRST;
Becker, White, \& Helfand 1995) which currently covers only a third of
our objects.}.  In addition to the above mentioned radio-loud quasars
we detect PC\,0027+0525 to be radio-loud and BRI\,1050-0000,
RX\,J1759.4+6638, and PC\,2331+0216 to be intermediate between the two
radio classes (see also McMahon et~al. 1994). All other quasars (but
two) have $R$ upper limits which designate them as being radio-quiet.
This result is in agreement with other radio and far-IR studies of
these $z>4$ quasars which find them to be radio-quiet (e.g., Schneider
et~al. 1992; Omont et~al. 1996; McMahon et~al. 1999). Our results are
also consistent with the Schmidt et~al. (1995) conclusion that only
5--10\% of optically selected $z>4$ quasars are radio-loud.

In Fig.~\ref{aoxhis} we compare the $\alpha_{\rm ox}$ distribution of
all the X-ray detected $z>4$ quasars with the $\alpha_{\rm ox}$
distribution of a sample of all 87 Palomar-Green (PG) quasars at
$z<0.5$ from Brandt, Laor, \& Wills (2000). We have translated the
$\alpha_{\rm ox}$ given in Brandt et~al. (2000) for a flux density at
3000~\AA\ to that for a flux density at 2500~\AA . To carry out this
comparison we consider only the optically selected $z>4$
quasars\footnote{The only objects which are not optically selected in
our study are the five last objects listed in Table~1.} since
the PG sample is an optically selected sample.  The $\alpha_{\rm ox}$
distribution for the seven optically-selected objects is consistent
with the $\alpha_{\rm ox}$ distribution of the PG quasars and with the
$\alpha_{\rm ox}$ distribution usually found for quasars (e.g., Wilkes
et al. 1994; Green et~al. 1995).  We also use all optically selected
objects in this study, including the ones with upper limits on their
X-ray properties, to derive the mean $\alpha_{\rm ox}$.  To that
purpose we have used the ASURV software package Rev 1.2 (LaValley,
Isobe \& Feigelson 1992), which implements the survival analysis
methods presented in Feigelson \& Nelson (1985) and Isobe, Feigelson,
\& Nelson (1986). 
%
\centerline{\epsfxsize=8.5cm\epsfbox{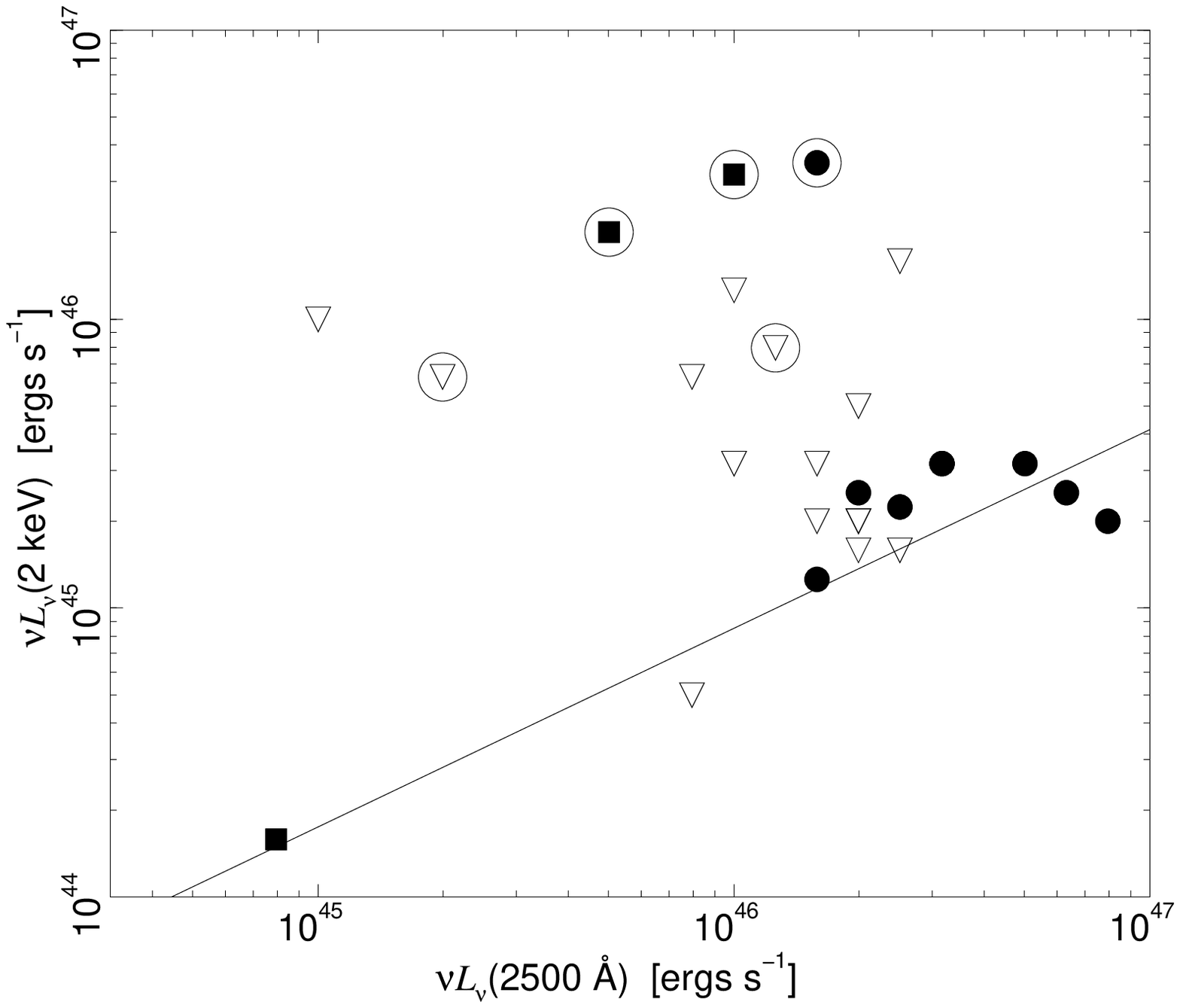}}
\figcaption{Rest-frame 2~keV versus 2500~\AA\ luminosities. Filled circles
are data from this study, squares are data from previous studies, and
empty triangles are upper limits determined in this study.  Radio-loud
quasars are circled.  The separation between the three blazar-type
(highest points) from the other objects is clearly visible.
\label{lxvslo}}
\centerline{}
\centerline{}

\noindent We find the mean  $\alpha_{\rm ox}$ to be $-1.49\pm
0.04$ which is in agreement with past studies.

Three of the detected quasars have $\alpha_{\rm ox} < -1$
(Fig.~\ref{aoxhis}). This is not consistent with the PG quasars'
$\alpha_{\rm ox}$ distribution, and indeed these three quasars are the
ones known to be radio-loud with two of them being blazar-type objects,
while the PG quasars are mainly radio-quiet with no blazars among them.

The separation between the blazar-type and radio-quiet quasars can also
be seen in Fig.~\ref{lxvslo}, where we plot the rest-frame 2~keV versus
2500~\AA\ luminosities.  The X-ray fluxes for the three radio-loud
quasars are about an order of magnitude higher than those of the
radio-quiet ones.  A line fit to the radio-quiet quasars' data using
ASURV yields
%
%
\begin{equation}
\frac{\nu L_\nu (2\,{\rm keV})}{10^{45}\,{\rm ergs\,s}^{-1}}=
\left(0.175^{+0.070}_{-0.050}\right)
\left[\frac{\nu L_\nu (2500\,{\mbox\AA})}{10^{45}\,{\rm ergs\,s}^{-1}}
\right]^{0.69\pm 0.10}   \ \ .
\end{equation}
This relation is in agreement with the one found for lower-redshift
quasars (e.g., Wilkes et al. 1994; Green et~al. 1995), albeit within
relatively large uncertainty owing to the small number of points and
their distribution.

Three of our detected quasars have two {\it ROSAT} observations
(Table~\ref{properties}). In all cases the fluxes from different
observations agree to within the measurement uncertainty of $\sim$30\%,
and no flux variations over time are detected.  The quasar
GB\,1428+4217 has six {\it ROSAT} observations and was found to vary by
a factor of two over a timescale of two weeks (or less), which
corresponds to less than 2.5 days in the source's rest-frame, a result
not unexpected in that this quasar is a flat-spectrum radio-loud blazar
(Fabian et~al. 1998, 1999).

The observed 0.1--2 keV band corresponds to a rest-frame band of
0.5--10 keV at $z=4$. Emission at the low end of this bandpass can
originate from  an accretion disk, but is mainly thought to arise from
the surrounding corona (e.g., Fabian 1994). Detecting X-ray emission in
this band from $z>4$ quasars suggests that similar processes are taking
place in low-redshift and the highest-redshift quasars.

We have established that in almost all cases where $z>4$ quasars were
observed at the center of the PSPC field and the exposure times were
sufficiently long (as anticipated from their optical luminosities), an
X-ray source was found at the optical position of the quasar. As two,
and possibly three, of the objects detected before our paper are
``peculiar'' blazar-type objects, we have more than doubled the number
of optically selected $z>4$ quasars detected in the X-ray band and
determined that $\alpha_{\rm ox}$ in these quasars is similar to that
of lower $z$ quasars. At present we have only been able to study in
X-rays the most luminous $z>4$ quasars. New X-ray missions such as {\it
Chandra}, {\it XMM}, {\it Constellation-X}, and {\it XEUS} should allow
the study of the X-ray properties of considerably less luminous $z>4$
quasars. With thousands of $z>4$ quasars expected to be found in the
next few years (e.g., by the Sloan Digital Sky Survey), there should be
ample targets.

\acknowledgments

We are grateful for several valuable suggestions by David J.
Helfand. We acknowledge the support of NASA LTSA grant NAG5-8107 (SK,
WNB), the Alfred P. Sloan Foundation (WNB), and NSF grant 99-00703
(DPS). We thank Harald Ebeling for the use of his IDL software.


\begin{thebibliography}{}
\bibitem{dum1} Arnaud, K. A. 1996, in Astronomical Data Analysis Software and Systems,
      ASP Conference Series 101, ed. G. Jacoby, \& J. Barnes,
     (San Francisco: ASP), 17
\bibitem{dum2} Bechtold, J., et al. 1994a, AJ, 108, 374
\bibitem{dum3} Bechtold, J., et al. 1994b, AJ, 108, 759
\bibitem{dum4} Becker, R. H., White, R. L., \& Helfand, D. J. 1995, AJ, 450, 559
\bibitem{dum5} Blair, A. J., Stewart, G. C., Georgantopoulos, I., Boyle, B. J., Almaini, 
      O., Shanks, T., Gunn, K. F., \& Griffiths, R. E. 1998, Astr. Nach., 
      319, 25
\bibitem{dum6} Brandt, W. N., Laor, A., \& Wills, B. J. 2000, ApJ, in press
      (astro-ph/9908016)
\bibitem{dum7} Condon, J. J., Cotton, W. D., Greisen, E. W., Yin, Q. F., Perley, 
      R. A., Taylor, G. B., \& Broderick, J. J. 1998, AJ, 115, 1693
\bibitem{dum8} David, L. P., Harnden, F. R., Kearns, K. E., \& Zombeck, M. V. 1999,
      The {\it ROSAT} High Resolution Imager Calibration Report.
      U.S. {\it ROSAT} Science Data Center, Cambridge
\bibitem{dum9} Dickey, J. M., \& Lockman, F. J. 1990, ARA\&A, 28, 215
\bibitem{dumq} Ebeling, H., White, D. A., \& Rangarajan, F. V. N. 2000, MNRAS, submitted
\bibitem{dumw} Efstathiou G., \& Rees M. J. 1988, MNRAS, 230, L5
\bibitem{dume} Elvis, M., Fiore, F., Giommi P., \& Padovani P. 1998, ApJ, 492, 91
\bibitem{dumr} Fabian, A.C. 1994, ApJS, 92, 555
\bibitem{dumt} Fabian, A. C., Brandt, W. N., McMahon, R. G., \& Hook, I. M.
      1997, MNRAS, 291, L5
\bibitem{dumy} Fabian, A. C., Iwasawa, K., Celotti, A., Brandt, W. N., 
      McMahon, R. G., \& Hook, I. M. 1998, MNRAS, 295, L25
\bibitem{dumu} Fabian, A. C., Celotti, A., Pooley, G., Iwasawa, K., Brandt, W. N., 
      McMahon, R. G., \& Hoenig, M. D. 1999, MNRAS, 1999, 308, L6
\bibitem{dumi} Fan, X., et al. 1999, AJ, 118, 1
\bibitem{dumo} Fan, X., et al. 2000, AJ, 119, in press
\bibitem{dump} Feigelson, E. D., \& Nelson, P. I. 1985, AJ, 293, 192
\bibitem{duma} Fiore, F., Elvis, M., Giommi, P., \& Padovani, P. 1998, ApJ, 492, 79
\bibitem{dums} Green, P. J., et al. 1995, ApJ, 450, 51
\bibitem{dumd} Hawkins, M. R. S., \& V\'{e}ron, P. 1996, MNRAS, 281, 348
\bibitem{dumf} Henry, J. P., et al. 1994, AJ, 107, 1270
\bibitem{dumg} Hook, I. M., \& McMahon, R. G. 1998, MNRAS, 294, L7
\bibitem{dumh} Isobe, T., Feigelson, E. D., \& Nelson P. I. 1986, AJ, 306, 490
\bibitem{dumj} Kellermann, K. I., Sramek, R., Schmidt, M., Shaffer D. B., Green, R. F.
      1989, AJ, 98, 1195
\bibitem{dumk} LaValley, M., Isobe, T., \& Feigelson, E. D. 1992, in Astronomical
      Data Analysis Software and Systems, ASP conference series 25,
      ed. D. M. Worrall, C. Biemesderfer, \& J.  Barnes, 
     (San Francisco: ASP), 245
\bibitem{duml} Mathur, S., \& Elvis, M. 1995, AJ, 110, 1551
\bibitem{dumz} McMahon, R. G., Omont, A., Bergeron, J., Kreysa, E., Haslam, C. G. T.
      1994, MNRAS, 267, L9
\bibitem{dumx} McMahon, R. G., Priddey, S. R., Omont, A., Snellen, I., \&
      Withington, S. 1999, MNRAS, 309, L1
\bibitem{dumc} Molthagen, K., Wendker, H. J., \& Briel, U. G. 1995, A\&A, 295, 43
\bibitem{dumv} Moran, E. C., \& Helfand, D. J. 1997, ApJ, 484, L95
\bibitem{dumb} Moran, E. C., Helfand, D. J., Becker, R. H., \& White, R. L.
      1996, ApJ, 461, 127
\bibitem{dumn} Mukai, K. 1997, The {\sc pimms} Users' Guide (Greenbelt: NASA/GSFC)
\bibitem{dumm} Netzer, H. 1990, in Active Galactic Nuclei, ed., T. J. -L. Courvoisier,
      \& M. Mayor (Berlin: Springer-Verlag), 57
\bibitem{dumqq} Omont, A., McMahon, R. G., Cox, P., Kreysa, E., Bergeron, J.,
      Pajot, F., \& Storrie-Lombardi, L. J. 1996, A\&A, 315, 1
\bibitem{dumww} Pfeffermann, E., et~al., 1987, Proc. SPIE, 733, 519
\bibitem{dum} Reeves, J. N., Turner, M. J. L., Ohashi, T., \& Kii, T. 1997, MNRAS,
      292, 468
\bibitem{dumee} Schmidt M., van Gorkom J. G., Schneider D. P., Gunn J. E. 1995, AJ, 
      109, 473
\bibitem{dumrr} Schneider, D. P. 1999, in After the Dark Ages: When Galaxies were Young, 
      ed. S. Holt, \& E. Smith, (American Institute of Physics Press), 233
\bibitem{dumtt} Schneider, D. P., Schmidt, M., \& Gunn J. E. 1989, AJ, 98, 1507
\bibitem{dumyy} Schneider, D. P., Schmidt, M., \& Gunn J. E. 1991, AJ, 101, 2004
\bibitem{dumuu} Schneider, D. P., van Gorkom, J. H., Schmidt, M., \& Gunn, J. E. 1992,
      AJ, 103, 1451
\bibitem{dumii} Schneider, D. P., Schmidt, M., Hasinger, G. Lehmann, I., Gunn, J. E.,
      Giacconi, R., Tr\"{u}mper, J., \& Zamorani, G. 1998, AJ, 115, 1230
\bibitem{dumoo} Shaver, P. A., Wall, J. V., \& Kellermann, K. I. 1996, MNRAS, 278, L11
\bibitem{dumpp} Storrie-Lombardi, L. J., McMahon, R. G., Irwin, M. J., \& Hazard, C.
      1996, ApJ, 468, 121
\bibitem{dumaa} Turner, E. L. 1991, AJ, 101, 5
\bibitem{dumss} Voges, W., et~al. 1996, in R\"{o}ntgenstrahlung from the Universe,
      ed. H. U. Zimmermann, J. Tr\"{u}mper, \& H. Yorke, MPE Report 263, 637
\bibitem{dumdd} Warren, S. J., Hewett, P. C., Irwin, M. J., McMahon, R. G., \& 
      Bridgeland, M. T. 1987, Nature, 325, 131
\bibitem{dumff} Wilkes, B. J, Tananbaum, H., Worrall, D. M., Avni, Y., Oey, M. S., \&
      Flanagan, J.  1994, ApJS, 92, 53
\bibitem{dumgg} York, D. G., et al. 2000, AJ, submitted
\bibitem{dumhh} Zickgraf, F. -J., Voges, W., Krautter, J., Thiering, I., 
      Appenzeller, I., Mujica, R., \& Serrano, A. 1997, A\&A, 323, L21
\end{thebibliography}
\end{document}